\documentclass{svproc}
\usepackage{color}
\usepackage{comment}

\mathchardef\mhyphen="2D

\usepackage[colorinlistoftodos]{todonotes}
\usepackage{lipsum}
\usepackage{amsmath}
\usepackage{amssymb}
\usepackage[ruled,vlined,linesnumbered,lined,boxed]{algorithm2e}

\newcommand{\ColorPathProblem}{{\sf Maximum Colorful Path in Temporal Graph}}

\newcommand{\ColorPath}{{\sf Max CPTG}}

\newcommand{\CTPLS}{{\sf Colorful Temporal Path Local Search}}

\newcommand{\TDom}{{\mathcal{T}}}
\newcommand{\IS}{{\sf Max IS}}

\usepackage{url}

\begin{document}
\mainmatter              % start of a contribution
\title{Finding Colorful Paths in Temporal Graphs}

%
%\titlerunning{Maximum Colorful Paths in Temporal Graphs}  % abbreviated title (for running head)
\author{Riccardo Dondi\inst{1} \and
Mohammad Mehdi Hosseinzadeh\inst{1}}
\authorrunning{Riccardo Dondi et al.} % abbreviated author list (for running head)
\tocauthor{Riccardo Dondi and Mohammad Mehdi Hosseinzadeh}
\institute{Universit\`a degli Studi di Bergamo, Bergamo, Italy,\\
\email{riccardo.dondi@unibg.it}; \email{m.hosseinzadeh@unibg.it}
}

\maketitle              % typeset the title of the contribution

\begin{abstract}
The problem of finding paths in temporal graphs 
has been recently considered due to its many 
applications.
In this paper we consider a variant of the problem 
that, given a vertex-colored temporal graph, 
asks for a path whose vertices
have distinct colors and include 
the maximum number of colors. 
We study the approximation
complexity of the problem and
we provide an inapproximability lower bound. 
Then we present a heuristic for the problem 
and an experimental evaluation of our heuristic,
both on synthetic and real-world graphs.

\keywords{Temporal Graphs, Algorithms on Graphs, 
Algorithms for Network Analysis, Heuristics, Approximation Complexity.}
\end{abstract}

\section{Introduction}
\label{section:intro}

Finding paths is a basic problem  in graph theory \cite{DBLP:books/daglib/0030488} and several variants
have been studied, including finding a shortest path between two  vertices and finding a longest path in a graph.
Recently, these problems have been considered for 
real-world data that need a description
of the vertex properties and dynamics of the relations \cite{DBLP:journals/bigdata/ThejaswiGL20}.
For these data, a richer representation with respect to the classical graph model has to be introduced, for example
by associating labels or colors with vertices and 
by representing the evolution of
relations with a temporal graph. In this latter model, 
edges are associated with timestamps to represent 
when an interaction occurred \cite{holme2015modern}.

In this paper we consider a problem that
looks for a path in a temporal graph that has vertices
associated with colors. 
Given a set of colors, the problem asks
for a temporal path having vertices 
with distinct colors and including the maximum number of colors. A temporal
path in a temporal graph is a path in which the timestamps of consecutive edges are strictly
increasing, thus representing a path
that does not violate the time constraint
specified by the timestamps of the edges.
The problem we consider is a variant of 
the one considered in \cite{DBLP:journals/bigdata/ThejaswiGL20},
that asks for a temporal path that exactly matches
a multiset of colors (called motif in \cite{DBLP:journals/bigdata/ThejaswiGL20}).
% \footnote{in \cite{DBLP:journals/bigdata/ThejaswiGL20} the motif can be a
% multiset is some cases; here we focus on the case that the motif is a set}). % introduced in ...
As outlined in \cite{DBLP:journals/bigdata/ThejaswiGL20},
this problem has several applications, for example in  
tour recommendations \cite{DBLP:conf/ht/ChoudhuryFAGLY10,DBLP:conf/wsdm/GionisLPT14}, where vertices correspond 
to interesting locations,  colors represent activities available in locations, edges correspond to transportation links between different locations (a timestamp is associated for example
to departure time).
A set (or a multiset) of colors represents  activities
a tourist may be interested into
and a path associated with different colors 
is then a suggestion of
the activities that can be carried out
respecting the time constraints.
% A second application, described in \cite{DBLP:journals/bigdata/ThejaswiGL20}, is in analysis of financial transactions. 
% Each vertex represents a financial entity and 
% colors represent their properties.
% Two entities are connected by a temporal edge
% when a financial transactions between them occurs. A financial analyst may be interested in 
% chains of transactions involving entities
% with distinct properties, for example
% money laundering activities may involve public figures, companies with certain types of contracts, and banks in offshore locations.
%Given a set of colors, 
%Due to the structure of a temporal network, %or the time 
%constraint,
A temporal graph, due to its structure, may not 
contain a temporal path that includes
all the colors. 
% it may not exist a temporal path in a temporal graph that includes
% all the colors. 
Thus a natural direction that we consider
in this paper is to look for a temporal path that includes the
maximum number of colors. 

%\subsubsection*{Related Works.}

\paragraph{Related Works.}
Given a (static) vertex-colored graph,
the problem of finding a colored path
whose vertices have distinct colors and that includes the maximum number of colors (called tropical path) has been recently investigated 
in \cite{TropicalPath}. 
%, a problem called tropical path. 
% where 
% it is considered the
% problem of finding a path in a graph that
% includes the maximum number of 
% colors. 
In \cite{TropicalPath}, it is shown that the problem is not
approximable, unless P = NP, 
within constant factor as the Longest
Path problem, while hardness results or polynomial-time algorithms are given
for several graph classes (bipartite chain graphs, threshold
graphs, trees, block graphs, and proper interval graphs).
A related problem on static graphs
is that of finding a path
whose vertices contains all
the colors in a set and
the vertices in the path are all
colored distinctly \cite{DBLP:journals/jacm/AlonYZ95,DBLP:journals/tcs/KowalikL16}.
% Another related problem considered in vertex-colored static graphs is that
% of finding the maximum number of vertex-disjoint uni-color paths \cite{DBLP:journals/jco/DondiS18,DBLP:journals/algorithms/BonizzoniDP13}.

Several variants of the problem
of finding a temporal path
in a temporal vertex-colored graph
that matches a given multiset of colors (called motif)
%in a colored temporal graph 
have been introduced in \cite{DBLP:journals/bigdata/ThejaswiGL20}.
% , where
% given a multiset of colors (called motif), 
% the  \cite{DBLP:journals/bigdata/ThejaswiGL20} considers
% several variants of the problem of finding a temporal path 
% that matches the motif.
% (the vertices of the path 
% have a multiset of colors defined by the motif).
% query;
% .
% Several decision variants of the problem have been considered, mainly asking if there exists a temporal
% path whose vertices contain distinct colors,
% or matching the given motif.
It is shown in \cite{DBLP:journals/bigdata/ThejaswiGL20}
that these variants of the problem are NP-complete,
but fixed-parameter tractable when parameterized
by the size of the motif.

Several problems related
to finding paths in a temporal graph have been considered \cite{DBLP:journals/pvldb/WuCHKLX14,DBLP:journals/tkde/WuCKHHW16}. 
A notable example is that of checking whether there exists 
a temporal path with
waiting time constraint, a problem
that has been shown to be NP-complete \cite{DBLP:conf/isaac/CasteigtsHMZ20}.
A similar problem is the temporal graph exploration 
\cite{DBLP:journals/jcss/Erlebach0K21}
that asks for a temporal walk that, starting 
at a given vertex, visits all vertices of a graph
with the smallest arrival time.
Other related problems ask
for the deletion of vertices so that temporal paths connecting pairs of vertices are 
removed \cite{DBLP:journals/jcss/ZschocheFMN20}.
Some recent contributions have investigated the
computational complexity of exploring
a temporal graph when the underlying graph is a star
and finding an eulerian walk in a temporal graph
%showing that the problem is NP-complete 
\cite{DBLP:journals/jcss/AkridaMSR21,DBLP:conf/iwoca/BumpusM21,DBLP:conf/iwoca/MarinoS21}.
%% AGGIUNGERE riferimenti IWOCA 2021

%\subsubsection*{Our Contribution.}
\paragraph{Our Contribution.}
In this paper, given a temporal
vertex-colored graph,
we consider the problem 
of finding a temporal path whose vertices
have distinct colors and that 
includes the maximum number of colors (a problem
called \ColorPath{}).
First, we study the approximation
complexity of the \ColorPath{} problem and
we show in Section \ref{section:inapprox}
that it is not approximable within factor $O(|V|^{\frac{1}{2}- \varepsilon})$,
unless $P = NP$. Notice that the corresponding 
problem on static graphs (finding a tropical path)
is only known to be not approximable
with constant factor, unless $P=NP$ \cite{TropicalPath}.

In Section \ref{sec:heuristic}
we present a heuristic for \ColorPath{},
as our aim is to design a method
that is applicable even for a large number of colors.
Notice that the methods proposed in \cite{DBLP:journals/bigdata/ThejaswiGL20} 
%are different
%from our contribution, since  
are for different variants of the problem, 
% are considered in\cite{DBLP:journals/bigdata/ThejaswiGL20}, 
where
all the colors of the motif have to be included in a solution. 
Moreover, the methods proposed in  \cite{DBLP:journals/bigdata/ThejaswiGL20}
are fixed-parameter algorithms, where the parameter is the size
of the motif, hence the running time
of these latter algorithms is exponential in the
size of the motif, leading to methods that are
able to process motifs of moderate size (up to $18$ colors are considered in  \cite{DBLP:journals/bigdata/ThejaswiGL20}).
On the other hand, we have to point out that
the methods in \cite{DBLP:journals/bigdata/ThejaswiGL20} compute
exact solutions, while our method is only a heuristic.
In Section \ref{sec:exp}, we present an
experimental evaluation of our heuristic,
both on synthetic and real-world graphs.
We start in Section \ref{section:preliminaries}
by introducing some definitions and by defining
the problem we are interested into.
Some proofs are omitted due to space constraints (marked by $\star$).

\section{Preliminaries}
\label{section:preliminaries}

We start this section by introducing the definition of
discrete time domain over which is defined 
a temporal graph.

\begin{definition}
A discrete time domain $\TDom$ is a sequence of
timestamp $t_i$, $1 \leq i \leq t_{max}$, where
each $t_i$ is an integer and $t_i < t_{i+1}$. 
An interval $T=[t_i,t_j]$ over
$\mathcal{T}$, 
where $t_i,t_j \in \mathcal{T}$ and $t_i \leq t_j$,
is the sequence of timestamps 
$t$ such that $t_i \leq t \leq t_j$.
%having values between $t_i$ and $t_j$,.
\end{definition}

Two intervals $T_1 =[t_{a,1}, t_{b,1}]$ and
$T_2 =[t_{a,2}, t_{b,2}]$
are disjoint if they do not share any timestamp,
that is
$t_{a,1} \leq t_{b,1} < t_{a,2} \leq t_{b,2}$
or
$t_{a,2} \leq t_{b,2} < t_{a,1} \leq t_{b,1}$.
The concatenation of $T_1$ and $T_2$ is
an interval $T_1 \cdot T_2$
obtained by merging the two time intervals $T_1$ and $T_2$,
that is, assuming without loss of generality
that $t_{a,1} \leq t_{b,1} < t_{a,2} \leq t_{b,2}$,
$
T_1 \cdot T_2  = [t_{a,1} , t_{b,2}].
$
Given a set of pairwise disjoint intervals
$T_1 =[t_{a,1}, t_{b,1}]$,
$T_2 =[t_{a,2}, t_{b,2}]$, \dots ,
$T_q =[t_{a,q}, t_{b,q}]$, where
$t_{a,1} \leq t_{b,1} < t_{a,2} \leq t_{b,2}
< \dots  < t_{a,q} \leq t_{b,q}$,
we can define the concatenation
of these intervals:
$
T_1 \cdot T_2 \cdot \dots T_q = [t_{a,1} , t_{q,2}].
$

We present now the definition of temporal graph. We assume
that the vertex set is not changing on the time domain,
that is the vertex set is identical in each timestamp. 

\begin{definition}
\label{def:TempGraph}
A temporal graph  $G = (V,E,\TDom)$ consists of

\begin{enumerate}

\item A set $V$ of vertices

\item A time domain $\TDom$

\item A set $E \subseteq V \times V \times \TDom$ of
temporal edges, where a temporal edge of $G$ is 
a triple $\{u,v,t\}$,
with $u,v \in V$ and $t \in \TDom$.

\end{enumerate}
\end{definition}

$E[t]$ denotes the set of active edges at timestamp $t \in \TDom$, that is:
$
E[t]=\{\{u,v,t\}|\{u,v,t\}\in E \}.
$
%We assume hence that all the nodes belongs to the overall
%time domain and 
%while $V[T$

Now, we introduce the definition of temporal path.

\begin{definition}
\label{def:TempPath}
Given a temporal graph $G=(V,E,\TDom)$, a temporal path in 
$G$ is 
an alternating sequence of vertices and temporal edges $v_{p,1}\ e_{p,1}\ v_{p,2}\ e_{p,2}  \dots \ e_{p,q-1}\ v_{p,q}$
such that:

\begin{enumerate}

\item $v_{p,1}$, $v_{p,2}$, $\dots$, $v_{p,q}$ are distinct
vertices

\item For each $i$, with $1 \leq i \leq q-1$,
$e_{p,i} = \{v_{p,i}, v_{p,i+1},t_i \} \in E$, with $t_i \in \TDom$

\item For each $i$, with $1 \leq i \leq q-1$,
it holds $t_i < t_{i+1}$.

\end{enumerate}

\end{definition}

Vertices $v_{p,1}$ and $v_{p,q}$ in $p$ are the start and end vertex of $p$. 
The length of $p$, denoted by $|p|$,
is the number of vertices in $p$.
We refer to Point 3 of Definition \ref{def:TempPath}
as the \emph{time constraint} of a temporal path.

A vertex-colored temporal graph is defined by adding 
a coloring to the vertices of 
a temporal graph.

\begin{definition}
$G_c = (V,E,\TDom,c)$ is a vertex-colored temporal graph,
where $G = (V,E,\TDom)$ is a temporal graph and 
$c: V \rightarrow C$ is a function that assigns a color from set $C$ to each vertex in $V$.

% consist of

% is defined as follows:

% \begin{itemize}

% \item $G = (V,E,\TDom)$ is a temporal graph

% \item $c: V \rightarrow C$ is a function that assigns a color from set $C$ to each vertex 
% in $V$.

% \end{itemize}
\end{definition}

We can now define the concept of colorful 
set of vertices.

\begin{definition}
Given a vertex-colored temporal graph $G_c = (V,E,\TDom,c)$, a set $V' \subseteq V$ is 
\emph{colorful} if 
all the vertices in $V'$ have distinct colors.
\end{definition}

A temporal path in a vertex-colored temporal graph  $G_c = (V,E,\TDom,c)$ is colorful if all its vertices have distinct colors.
% \begin{definition}
% Given a colored temporal graph $G_c = (V,\TDom,E,c)$, 
% a colorful temporal path $p =[v_{p,1}, t_1 v_{p,2}, t_2 \dots, v_{p,q}]$
% is a temporal path in $G_c$ such that the set
% $\{v_{p,1}, v_{p,2}, \dots, v_{p,q}\}$ is colorful, that is all the vertices have distinct colors.
% \end{definition}
Now, we are able to define the problem we are interested
into.

\begin{problem}\ColorPathProblem{} (\ColorPath) \\ 
\textbf{Input:} A vertex-colored temporal colored graph $G = (V,\TDom,E,c)$.\\
\textbf{Output:} A colorful temporal path in $G$ 
that includes the maximum number of colors
(that is it has maximum length).\\
\end{problem}

\section{Inapproximability of \ColorPath}
\label{section:inapprox}

In this section we prove that the \ColorPath{} problem 
cannot be approximated within factor 
$O(|V|^{\frac{1}{2} - \varepsilon})$, unless $P= NP$.
%or $O(|\sqrt{\mathcal{T}}|)$.
We prove this result by giving an approximation
preserving reduction from the {\sf Maximum Independent Set} problem
(denoted by \IS). 
For details on approximation
preserving reductions see
\cite{DBLP:books/daglib/0030297}.
The \IS{} problem, given a graph
$G_I = (V_I,E_I)$, where $|V_I| = n$ and $|E_I| = m$,
asks for an independent set $I \subseteq V_I$ of maximum size
(we recall that $I$ is an independent
set if for $u,v \in I$, it holds that  $\{u,v\} \notin E_I$).

% We present here the definition of the \IS{} problem.

% \begin{problem}\IS\\ 
% \textbf{Input:} A graph $G_I = (V_I,E_I)$, where
% $|V_I| = n$ and $|E_I| = m$.\\
% \textbf{Output:} An independent set $I \subseteq V_I$ of maximum size.\\
% \end{problem}

%%%%%%%%%%%%%%%%%%%%%%%%%%%%%%%%%%%%%%%
% TODO: Sostituire n+1 con z? %%%%%%%%%
%%%%%%%%%%%%%%%%%%%%%%%%%%%%%%%%%%%%%%%

\begin{figure}
\centering
    \includegraphics[scale=0.425]{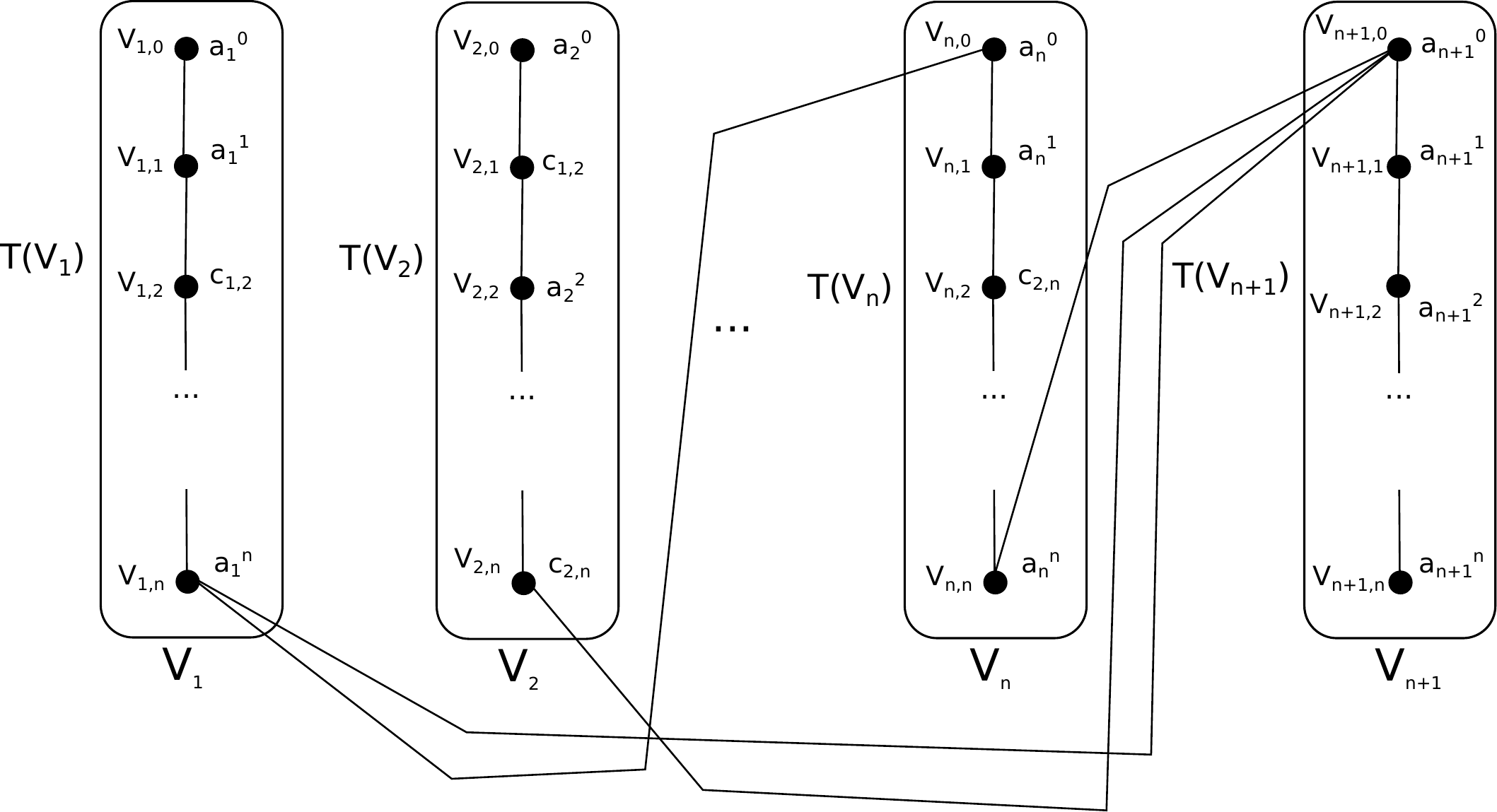}
    \caption{An overview of the temporal
    graph $G=(V,\TDom, E,c)$ associated with $G_I$.
    Each box contains the set $V_i$ of vertices,
    $1 \leq i \leq n+1$, and the path $p(V_i)$; the temporal edges are active in time interval $T(V_i)$ (on the left of the box).
    Each vertex is labeled on the left with its name, on the right with its color. We assume that
    $\{ v_1, v_2 \} \in E_I$, hence $c(v_{1,2})=c(v_{2,1})= c_{1,2}$, 
    $\{ v_1, v_n \} \notin E_I$, hence $c(v_{1,n}) = a_1^n$, $c(v_{n,1}) = a_n^1$
    and $\{ v_2, v_n \} \in E_I$, hence $c(v_{2,n})=c(v_{n,2})= c_{2,n}$.}
    \label{fig:Example}
\end{figure}

Next, we describe our approximation preserving reduction 
from \IS{} to \ColorPath{}.
Given an instance $G_I=(V_I, E_I)$ of \IS, 
we define a corresponding vertex-colored 
temporal graph $G=(V,\TDom, E,c)$, which is an 
instance of \ColorPath{} (an overview
of $G=(V,\TDom, E,c)$ is given in
Fig. \ref{fig:Example}).

%\begin{itemize}

For each $v_i \in V_I$, $1 \leq i \leq n$,
$V$ contains a set $V_i$ of
$n+1$ vertices:
$
V_i = \{ v_{i,x}: 0 \leq x \leq n  \}.
$
Furthermore, $V$ contains an additional set of vertices
$
V_{n+1} = \{ v_{n+1,x}: 0 \leq x \leq n  \}.
$
The vertex set $V$ of $G$ is defined as follows:
$
V = \bigcup_{i=1}^{n+1} V_i$.

The time domain $\TDom$ consists of the 
concatenation of 
$n+1$ time intervals  $T(V_1), \dots, T(V_n), T(V_{n+1})$, where each $T(V_i)$, $1 \leq i \leq n+1$,
is associated with vertex set $V_i$.
The idea is that 
only edges connecting
vertices of $V_i$
are active in interval $T(V_i)$, except for the last timestamp.
The interval $T(V_i)$, $1 \leq i \leq n+1$,
is defined as follows:

\[
T(V_i) = [ (n(i-1)+i, (n+1)i].
\]

Notice, for example, that
$T(V_1) =[1, n+1]$ and
$T(V_2) =[n+2, 2n+2]$ and so on.
By construction, the intervals $T(V_i)$, $1 \leq i \leq n+1$, are disjoint.
% as for each $i$ with $1 \leq i \leq n$,
% the first timestamp of interval $T(V_i)$ is
% $r_i = (n+1)i$ while the last timestamp  of 
% interval $T(V_{i+1})$ is $l_{i+1} = (n)(i)+i+1 = i(n+1)+1$, thus $l_{i+1} = r_i +1$.
The time domain $\TDom$ is then the concatenation of intervals $T(V_1), T(V_2), \dots, T(V_{n+1})$, that is
$ 
\TDom = T(V_1)\cdot T(V_2) \cdot  \ldots \cdot T(V_{n})\cdot T(V_{n+1}).
$

The color function $c : V \rightarrow C$, is defined over
the following set $C$ of colors:
$
C = \{ c_{i,0} : 1 \leq i \leq n+1  \} 
\cup 
\{ c_{i,j}: \{  v_i, v_j \} \in E_I \wedge i < j\}
\cup 
\{ a_i^q: 1 \leq i,q \leq n+1 \}
%\{ ca_{i,j}:  \in E_I\}
.
$

Essentially, each color $c_{i,j}$ encodes an edge $\{v_i,v_j\} \in E$, with $1 \leq i < j \leq n$,
each color $a_i^q$, $1 \leq q \leq n$, 
encodes the fact that $v_i$ is not adjacent
to vertex $v_q$. Notice that 
$a_i^q \neq a^q_i$.

Now, we define the function $c$.
% in particular the colors assigned to vertices
% in each $V_i$, $1 \leq i \leq n+1$.
For the vertices  in $V_i$, with $1 \leq i \leq n$, $c$ is defined as follows:

\begin{itemize}

\item $c(v_{i,0}) = a_i^0$

\item $c(v_{i,x}) = c_{i,x}$, 
if $\{v_i,v_x\} \in E$ and $1 \leq i < x \leq n$

\item $c(v_{i,x}) = c_{x,i}$, 
if $\{v_i,v_x\} \in E$ and $1 \leq x < i \leq n$

\item $c(v_{i,x}) = a_i^x$, 
$1 \leq i,x \leq n$, if $\{v_i,v_x\} \notin E$

%\end{itemize}

Notice that $c(v_{i,i}) = a_i^i$, for each $i$ with $0 \leq i \leq n$, as we assume that $G_I$ does not contain self loops.

\end{itemize}

For the vertices of $V_{n+1}$,
the function $c$ is defined as follows:
%\begin{itemize}
%\item 
$c(v_{n+1,x}) = a_{n+1,x}$,  $0 \leq x \leq n$.

%\end{itemize}

Next, we define the set of temporal edges of $G$.
For each time interval $T(V_i)$, $1 \leq i \leq n+1$,
$G$ contains a colorful temporal path 
$p(V_i)$ 
induced by the vertices $v_{i,x}$ with $0 \leq x \leq n$.
The temporal edges active in interval $T(V_i)$ are defined as follows.
% The colorful path $p(V_i)$ is induced by the
% following edges:
%(except for time stamp $n j$):
% \begin{itemize}
% \item 
At timestamp $t = n(i- 1) + i + x $, 
with $0 \leq x \leq n-1$, 
$\{ v_{i,  x} , v_{ i, x+1}, t \} \in E$;
notice that $\{ v_{i,  x} , v_{ i, x+1}, t \}$ is the only active  temporal edge of $G$ at timestamp~$t$.

%\end{itemize}

The temporal path $p(V_i)$ resulting from these edge is then:

\[
p(V_i) = v_{i,0}\ \{ v_{i,0}, v_{i,1}, n(i-1)+i \}\ v_{i,1} \ \{ v_{i,1}, v_{i,2}, n(i-1)+i+1\} \ \dots 
\]
\[
\dots \{ v_{i,n-1}, v_{i,n}, ni+i -1 \} \ v_{i,n}
\]

Notice that, since by construction
two intervals $T(V_i)$, $T(V_j)$, $1 \leq i < j \leq n$,
are disjoint, the colorful temporal paths $p(V_i)$,
$p(V_j)$ are active in disjoint
intervals. 

The set $E$ contains also temporal edges 
defined to connect temporal colorful
paths $p(V_i)$, $1 \leq i \leq n$. 
At timestamp $t = (n+1)i$, $1 \leq i \leq n$,
the  following temporal edges belong
to $E$:

\begin{itemize}

\item $\{ v_{i,n}, v_{z, 0}, t \}$, with 
$1 \leq i < z \leq n$, such that 
edge $\{ v_i, v_z \} \notin E_I$

\item $\{ v_{i,n}, v_{n+1, 0},t \}$

\end{itemize}

This completes the definition of the 
vertex-colored temporal graph 
$G=(V,E, \TDom,c)$.
We prove now a property of $G$.

% At time stamp $t = (n+1)i$, the
% following set of temporal edges are active and connect 
% path $p(V_i)$, $1 \leq i \leq n$, with  
% colorful paths $p(V_z)$, such that $\{v_i,v_z \}
% \notin E_I$ and with $p(V_{n+1})$:

%%%%%%%%%%%%%%%%%%%%%%%%%%%%%%%%%%%%%%%%%%%%%%%%%%%%%%%%%%%
%% TO DO: show that in interval [n(j-1)+j, (n+1)j]
%% we have a colorful path
%% vertex-colored graph
%%%%%%%%%%%%%%%%%%%%%%%%%%%%%%%%%%%%%%%%%%%%%%%%%%%%%%%%%%%

\begin{lemma}
\label{lem:IndColoring1}$(\star)$
Let $G_I=(V_I, E_I)$ be an instance of \IS{} and 
let $G=(V, \TDom, E, c)$ be the corresponding
instance of \ColorPath{}. Then:
\begin{enumerate}

\item Each temporal path $p(V_i)$, with $1 \leq i \leq n+1$, is colorful

\item The vertices in temporal paths $p(V_i)$, $p(V_j)$, with 
$1 \leq i < j \leq n$ and $\{ v_i,v_j\} \notin E_I$, 
have different colors.
\end{enumerate}
\end{lemma}

%%%%%%%%%%%%%%%%%%%%%%%%%%%%%%%%%%%%%%%%%%%%%%%%%%%%%%%%%%%
%%%%%%%%%%%%%%%%%%%%%%%%%%%%%%%%%%%%%%%%%%%%%%%%%%%%%%%%%%%
%%%%%%%%%%%%%%%%%%%%%%%%%%%%%%%%%%%%%%%%%%%%%%%%%%%%%%%%%%%

Now,  we show how to construct in polynomial time a solution
of \ColorPath{} from a solution of \IS{}.

\begin{lemma}
\label{lem:IndColoring2}
Let $G_I=(V_I, E_I)$ be an instance of \IS{} and 
let $G=(V, \TDom, E, c)$ be the corresponding
instance of \ColorPath{}.
Given a solution $I \subseteq V_I$
of \IS{}, we can construct in polynomial time
a solution of \ColorPath{} of length
at least $(|I|+1)(n+1)$.
\end{lemma}
\begin{proof}
Consider an independent set $I = \{ v_{i,1}, v_{i,2},  \dots, v_{i,b}\}$ of $V_I$, where $i_1 < i_2< \dots < i_b$.
Then define a solution $p$ of \ColorPath{} as follows. The temporal path $p$ includes the colored temporal paths 
$p(V_{i_x})$ in interval $T(V_{i_x})$,
$1 \leq x \leq b$,
and the temporal 
colored path $p(V_{n+1})$  in interval $T(V_{n+1})$.
These colored paths are connected in $p$ 
by the following temporal edges:
$p(V_{i_x})$ and $p(V_{i_{x+1}})$, $1 \leq x \leq b-1$
are connected by temporal edge 
$\{ v_{i_x,n}, v_{i_{x+1},0}, t\}$ with $t = (n+1) i_x$;
$p(V_{i_b})$ and $p(V_{n+1})$
are connected by temporal edge 
$\{ v_{i_b,n}, v_{n+1,0}, t\}$ with $t = (n+1) i_b$.

%Notice that, 
Since $v_{i,x}, v_{i,y} \in V_I$, with $1 \leq x < y \leq b$, are 
not adjacent in $G_I$, it follows from 
Lemma \ref{lem:IndColoring1}
that the vertices in $p(V_{i_x})$ and $p(V_{i_y})$
do not share any color. 
Since each vertex in 
$p(V_{n+1})$ has a color distinct from the other
vertices in $V$, it follows that $p$ is colorful.
Furthermore, notice that by construction, 
since the paths $p(V_i)$, $1 \leq i \leq n+1$,
are defined over disjoint intervals,
$p$
is a temporal path. Finally, notice that
$p$ consists of $|I|+1$ paths $p(V_i)$ each
of length $n+1$,
thus concluding the proof.
\qed
\end{proof}

%Now, we show how to construct in polynomial time 
A solution of \IS{} can be computed in polynomial time starting from a solution of \ColorPath{}.

\begin{lemma}
\label{lem:IndColoring3}$(\star)$
Let $G_I=(V_I, E_I)$ be an instance of \IS{} and let $G=(V, \TDom, E, c)$ be the corresponding
instance of \ColorPath{}. Given a solution of \ColorPath{} of length
$(q+1)(n+1)$, we can construct in polynomial time
an independent set
of $G_I$ of size at least $q$.
\end{lemma}

The inapproximability
of \ColorPath{} follow from Lemma \ref{lem:IndColoring2}, Lemma \ref{lem:IndColoring3} and from the inapproximability of \IS{} \cite{DBLP:conf/stoc/Zuckerman06}.

\begin{theorem}
\label{teo:inapprox}$(\star)$
\ColorPath{} is not approximable within factor $O(|V|^{1/2- \varepsilon})$ unless P = NP.
\end{theorem}

\section{A Heuristic for \ColorPath{}}
\label{sec:heuristic}

In this section, we present our efficient heuristic, called \CTPLS{} (CTPLS), for \ColorPath{} problem.  CTPLS consists of two phases: 
%
%\begin{itemize}
%\item 
(1) A greedy preliminary step that computes an 
initial solution, 
%\item 
(2) A local search step that looks for a possible improvement of the solution.
%\end{itemize}

We start by describing the preliminary greedy step. 
Given a vertex-colored temporal graph $G_c = (V,E,\TDom,c)$, 
first the step  
computes a segmentation of the time domain $\mathcal{T}$ in
$|C|$ %\alpha$ (a parameter) 
disjoint intervals of
equal length. 
%\todo{Riccardo: I'm not sure we should use $\alpha$, it is 
%never considered in the experimental part.}
% Notice $\epsilon$ number of timestamps where $\epsilon=\frac{t_{max}}{\alpha}$, that is the overall number of timestamps divided by a parameter $\alpha >0$. 
Then it greedily looks for
a temporal edge to be added to the path $p$
computed so far in each interval. 
The path $p$ is initialized 
as a
%First, It select an initial 
temporal edge in the first interval.
In the next intervals, the greedy step
%starting from the second, 
looks for a temporal edge
that connects the last vertex of $p$ to 
a vertex $v$ that is not included in $p$.
% , called a candidate vertex, such that
% $v$ is not in $p$ and it has distinct colors from the vertices already in $p$.
% When different candidate vertices exist,
% the vertex chosen among the candidates is the one
% having larger degree in the next interval,
% considering only vertices with colors not in $p$.

% node has been chosen in a way that the connected nodes to that should have different color, that means if we have more than one connected node with the candidate node with same color at the next interval, we count only one in the degree. We count only nodes that have different color than the colors that already have been considered in the path.

% for the path $p$ in order to compute the first solution.
% For each interval the CTPLS 

% compute a vertex in an interval that has a larger degree in the next interval. The candidate node as a larger degree node has been chosen in a way that the connected nodes to that should have different color, that means if we have more than one connected node with the candidate node with same color at the next interval, we count only one in the degree. We count only nodes that have different color than the colors that already have been considered in the path. 

%...(TO BE Written: special case of the first interval)...\\
%For Mehdi: I've rewritten the part above, check it is fine, also the first case

%If the path $p$ computed by the greedy preliminary step doesn't cover all the colors, 
Then CTPLS applies a local-search strategy, consisting of two different 
possible modifications of $p$ (unless
$p$ contains all the colors).
% following levels in order to possibly improve computed solution of the preliminary step of \ColorPath{} problem. 

\begin{enumerate}

\item {\bf LS1} (Edge replacement):
starting from the first edge of $p$, a
temporal edge $\{u,v,t\}$ is possibly replaced with
two temporal edges $\{u,x,t_1\}$,
$\{x,v, t_2\}$, with $t_1 < t_2$; 
notice that %the resulting temporal path
%can be computed if 
vertex $x$ is not already in $p$ and it must be colored 
differently from the vertices already 
in $p$; 
furthermore, all the temporal edges of
the new path
must satisfy the time constraint.

% and looks for another node $x$ with different color that not exist in the solution. The new node $x$ should be connected to the $p(v_i)$ and $p(v_{i+1})$ such that $x$ should be connected to $p(v_i)$ on the timestamp before be connected to $p(v_{i+1})$. 

% Then, if the solution $p$ from first level of local search doesn't cover all the colors the CTPLS uses the second level of the local search strategy in order to possibly improve computed solution of \ColorPath{} problem.

\item {\bf LS2} (Vertex replacement):
starting from the first vertex in the solution, it possibly replaces a vertex $x$ in $p$ and the temporal 
edges %$\{u,x,t_1\}$, $\{x,v,t_2\}$ 
incident in $x$,
with two vertices $y$ and $z$
and three temporal edges
so that %the temporal edges of
the new path
satisfies the time constraint.
%$\{u,y,t'_1\}$, $\{y,z,t'_2\}$, 
%$\{z,v,t'_3\}$, such that $t'_1 < t'_ 2 < t'_3$.
Notice that $y$ and $z$ must not be in $p$
and must have different
colors from the vertices of $p$ (except for the replaced
vertex $x$).
% and the three temporal edges added must satisfy the time constraint.

% delete a node $p(v_i)$ and looks for two other nodes with different colors that already have been considered in the path except of the color of the node $p(v_i)$ that has been deleted. The two new nodes should be connected to each others and one of them also should be connected to the $p(v_{i+1})$.
% When the vertex replacement improves a solution,
% thus it computes a colorful temporal path $p'$ starting
% from $p$, 
% If the solution has been improved we consider it as a new solution and in this case we have two following ways:

% I) The colors of both nodes are different that the color of $p(v_i)$: in this case we start applying second level of local search from the beginning of the new solution.

% II) The color of one of the added nodes is same as the color of $p(v_i)$: in this case we continue the second level of local search from the node $p(v_{i+1})$.

\end{enumerate}

\section{Experimental Results}
\label{sec:exp}

%To be written
%\todo{Mehdi, starts to describe the synthetic datasets}

In this section, we present an experimental evaluation of CTPLS on synthetic and real networks.
The CTPLS heuristic described in Section \ref{sec:heuristic} is implemented in  Python 3.7 using the NetworkX package for managing networks \cite{hagberg2008exploring}. We perform the experiments on MacBook-Pro (OS version 11.4) with processor 2.9 GHzIntel Core i5 and 8GB 2133 MHz LPDDR3 of RAM.
%, Intel Iris Graphics 550 1536 MB.

\subsubsection*{Synthetic Networks.}

In the first part of our experimental evaluation, we analyse the performance of CTPLS %to find a solution for the \ColorPath{} problem 
on synthetic datasets.
We start by describing the synthetic datasets, then
we discuss the results of CTPLS.

\textbf{Datasets.} 
% \todo{Mehdi, describe better how the graphs
%  are built: the path and then other
%  added vertices, etc..}
Each synthetic graph is built as follows.
First, we generate a temporal graph 
consisting of $500$ vertices 
over $90$ timestamps,
such that the topology of the graph is based on one
of the following models:
Erd{\"o}s-Renyi (ER) with parameter $p=0.1$,
Erd{\"o}s-Renyi with parameter $p=0.4$
and Barabasi-Albert (BA) with parameter equal to $10$. 
$|C|$ vertices of the graph are then chosen randomly, assigned a distinct colors
and it is defined a temporal path that
connects them. This ensures that 
each synthetic graph contains an optimal solution
including all the colors in $C$,
thus allowing to
compare the solutions returned by CTPLS
with an optimal one.
Then each of the remaining vertices of the graph is
assigned uniformly random colors from $C$. % with equal probability. 
We consider  the following sizes of $C$: 10, 20, 30 and 50 colors.
% It includes a colorful temporal path that
% contains all the colors in $C$.
% (that is including all the colors of set $C$) such that each edge of the optimal solution belongs to one timestamp over $\mathcal{T}$.
%Then, we add to this path, 
For each graph model and for each
size of $C$ considered, we generated 20 
independent synthetic graphs.
%instances.
% Therefore, we have three groups of generated graphs and each of them consist 20 instances such that for each instance exists at least an optional solution. For each generated graph we consider . 

\textbf{Outcome.} We present in Table \ref{table:synth} the results of our experimental evaluation on the synthetic datasets. In particular, we report
the minimum, maximum, average and standard deviation of the returned solutions
of CTPLS over $20$ instances
%independent synthetic graphs 
for each color set and each graph model.
Furthermore, we report the average running time (in seconds).
% Recall than the synthetic graphs are designed
% so that they always contains a path
% including all the colors of set $C$.
% This allows us to compare the quality of 
% solutions returned by CTPLS with optimal
% solutions (having length $|C|$).

As reported in Table \ref{table:synth}, the performances
of CTPLS degrade with the increasing
of the number of colors.
For the BA-based graphs, for example, 
for a set of $10$ colors
the returned solutions contains on average 
at least $91\%$ of the
colors in $C$, for $50$ colors the average 
number of colors contained in the 
returned solutions 
is $17.2$ out of $50$.
%$\frac{21}{50}$.
The experimental results show also that
the performances of CTPLS depend 
on the specific graph models.
For the ER model with $p=0.4$, the solutions 
returned are within $84\%$ of the optimal
solutions (for $50$ colors).
The performances are worse on ER with $p=0.1$,
within $28.6\%$ of the optimal
solution (for $50$ colors).
%and for the BA model. 
For the BA model,
the solutions returned by CTPLS are close
to the optimum only for the case of $10$ colors (within $91\%$ of the optimal
solution) and are on average $83.75\%$, $47.5\%$
and $34.4\%$ for $20$, $30$ and $50$ colors, respectively.
%However, 
It has to be pointed out that
the \ColorPath{} problem is hard to approximate, 
as shown
in Section \ref{section:inapprox},
so it is not surprising that
for some datasets the lengths of the solutions returned
by CTPLS are not close to the optimum.

The method is always fast on synthetic datasets, requiring at most $0.68$ seconds average
running time (ER model with $p=0.4$ 
and $50$ colors).

 \begin{table}[htb]
 \centering % centering table
 \caption {Performance of CTPLS on synthetic datasets, varying colors from 10 to 50.  We report
 %and the value of returned solutions (path) for the \ColorPath{} problem is 
 minimum, maximum, average and standard deviation over 20 independent synthetic networks for each different color set. The average running time is in seconds.}

 \begin{tabular}{|l||c|c||c|c||c|c||c|c||} % creating  columns
 \hline 
 \multicolumn{1}{|c|}{}&\multicolumn{2}{|c|}{color 10}&\multicolumn{2}{|c|}{color 20}&\multicolumn{2}{|c|}{color 30}&\multicolumn{2}{|c|}{color 50} \\
\hline
 BA&path& time&path& time&path& time&path&time\\ 
 \hline
Min&8 &- &10 &- &12 &- &14 &-\\ 
Max&10 &- &16 &- &17 &- &21 &-\\ 
Average&9.1 &0.06 &13.1 &0.08 &14.25 &0.11 &17.2 &0.13\\
SD&0.79 &- &1.41 &- &1.65 &- &2.28 &-\\
 \hline
 \hline
 \multicolumn{1}{|c|}{}&\multicolumn{2}{|c|}{color 10}&\multicolumn{2}{|c|}{color 20}&\multicolumn{2}{|c|}{color 30}&\multicolumn{2}{|c|}{color 50} \\
\hline
ER $p=0.1$&path& time&path& time&path& time&path&time\\ 
 \hline
Min&9 &- &11 &- &9 &- &5 &-\\ 
Max&10 &- &19 &- &25 &- &30 &-\\
Average&9.85 &0.09 &16.75 &0.15 &18.45 &0.14 &14.3 &0.11\\
SD&0.37 &- &1.86 &- &4.67 &- &8.35 &-\\
\hline
\hline
 \multicolumn{1}{|c|}{}&\multicolumn{2}{|c|}{color 10}&\multicolumn{2}{|c|}{color 20}&\multicolumn{2}{|c|}{color 30}&\multicolumn{2}{|c|}{color 50} \\
\hline
ER $p=0.4$&path& time&path& time&path& time&path&time\\  
 \hline
Min&10 &- &19 &- &25 &- &38 &-\\
Max&10 &- &20 &- &30 &- &46 &-\\ 
Average&10 & 0.24&19.8 &0.35 &28.3 &0.66 &42.4 &0.68\\
SD&0 &- &0.41 &- &1.17 &- &1.82 &-\\

 \hline % inserts single-line
\end{tabular}
% %}
 \label{table:synth}
 \end{table}

\subsubsection*{Real Networks.}

In the second part of our experimental evaluation, we analyse the performance of CTPLS %to find a solution for the \ColorPath{} problem 
on four real-world datasets.

\textbf{Datasets.} 
We consider four different real-world temporal graphs taken from SNAP \cite{snapnets} for testing CTPLS: 
College messages\footnote{\url{http://snap.stanford.edu/data/CollegeMsg.html}} (\emph{CollegeMsg}), Email EU core\footnote{\url{http://snap.stanford.edu/data/email-Eu-core-temporal.html}} (\emph{email-Eu-core-temporal}), Bitcoin alpha\footnote{\url{http://snap.stanford.edu/data/soc-sign-bitcoin-alpha.html}} (\emph{soc-sign-bitcoinalpha}) and  
Bitcoin otc\footnote{\url{http://snap.stanford.edu/data/soc-sign-bitcoin-otc.html}} (\emph{soc-sign-bitcoinotc}).
These temporal graphs are not colored,
hence, following the same approach of \cite{DBLP:journals/bigdata/ThejaswiGL20},
we assigned uniformly random colors from 
a set of $30$ colors and from a set of $50$ colors. 
We consider two variants for each of this network,
since the length of an optimal solution of \ColorPath{}
on these graph is unknown. 
Hence, in order to evaluate the results of CTPLS,
for each real-world temporal graph 
we consider the original graph 
(denoted by NO-OP) and a modified 
temporal graph, called YES-OP, obtained by adding 
a temporal colorful path that contains each colors in $C$.
This latter temporal graph contains an optimal solution
of length $|C|$. 

% We consider the variant of YES-OP in order to know at least exist an optimal solution, hence, we can evaluate quality of the solutions obtain by CTPLS. \\
% TODO Write Why, few modifications, 
% and write before NO-OP

The first dataset, \emph{CollegeMsg}, is taken from private messages sent on an online social network at the University of California, Irvine, where temporal edges 
represent private messages sent between users at a given time.
%. Vertices represent users and 
%a temporal edge $\{u, v, t\}$ represents  a private message sent by user $u$ to user $v$ at time $t$. 
The dataset contains 59835 temporal interactions, 1899 vertices and time domain $\mathcal{T}$ of length $|\mathcal{T}|=58911$.
The \emph{email-Eu-core-temporal} dataset
is generated based on incoming and outgoing emails between members of a large European research institution, 
where temporal edges 
represent emails  sent between users at a given time. %Vertices represent institution members and
%A temporal edge $\{u, v, t\}$ represents 
%an e-mail sent from user $u$ to user $v$
%that member $u$ sent an e-mail to member $v$ 
%at time $t$. 
This dataset contains 332334 temporal interactions, 986 vertices and time domain $\mathcal{T}$ of length $|\mathcal{T}|=207880$.
\emph{soc-sign-bitcoinalpha} and \emph{soc-sign-bitcoinotc}
are % who-trusts-whom
datasets of members who trade using Bitcoin on platforms called Bitcoin Alpha and Bitcoin OTC, respectively, to prevent transactions with risky users. %Vertices represent members and 
%A temporal edge represent rates given by one mem
A temporal edge $\{u, v, t\}$ represents
a rate of member $v$ given by member $u$ at time $t$.
\emph{soc-sign-bitcoinalpha} contains 24186 temporal interactions, 3783 vertices and time domain $\mathcal{T}$ of length $|\mathcal{T}|=1647$,
\emph{soc-sign-bitcoinotc} contains 35592 temporal interactions, 5881 vertices and time domain $\mathcal{T}$ of length $|\mathcal{T}|=35445$.
% \emph{soc-sign-bitcoinotc}
% is another %who-trusts-whom 
% dataset of members who trade using Bitcoin on a platform called  Bitcoin OTC to prevent transactions with risky users. %Vertices represent members and 
% A temporal edge $\{u, v, t\}$ represents a rate of member $v$ given by member $u$ at time $t$.
% This network contains 35592 temporal interactions, 5881 vertices and time domain $\mathcal{T}$ of length $|\mathcal{T}|=35445$.

\textbf{Outcome.}
In Table \ref{table:real} we report the 
number of colors included in the  solutions returned by CTPLS and the running time (in minutes) for the two groups of real datasets we considered (NO-OP and YES-OP).
As shown in Table \ref{table:real}, for the NO-OP networks with $30$ colors, CTPLS found in the worst case a path
containing $20$ out of $30$ colors
(\emph{soc-sign-bitcoinalpha}) and in the best case an optimal solition (\emph{email-Eu-core-temporal}). For the other two networks,
\emph{CollegeMsg} and \emph{soc-sign-bitcoinotc} networks, 
CTPLS found suboptimal solutions that contains a significative
number of colors, $27$ and $25$ colors out of $30$, respectively.

For the YES-OP networks with $30$ colors, we don't report the result for
\emph{email-Eu-core-temporal}, as CTPLS was able to find an optimal solution for this dataset in NO-OP network. 
The results are not significantly different from
the corresponding NO-OP datasets.
%For the other YES-OP networks, 
CTPLS found in one case,
the \emph{CollegeMsg}, a path with the same number of colors as 
for the corresponding NO-OP network.
In one case,  (\emph{soc-sign-bitcoinotc}) CTPLS found a larger number of colors ($27$ instead of $25$ out of $30$),
in another case (\emph{soc-sign-bitcoinalpha}) 
CTPLS found a slightly smaller number of colors
($19$ instead of $20$ out of $30$ colors).
This decreasing is due to the fact that
CTPLS considers a temporal edge that belongs to the YES-OP instance and not to the NO-OP 
instance and this 
prevents CTPLS to include all the
vertices of the solution of the NO-OP instance.

For the NO-OP networks with $50$ colors, CTPLS found in the worst case a path containing $36$ colors (\emph{soc-sign-bitcoinalpha}) and in the best case  (\emph{email-Eu-core-temporal}) 
$49$ out of $50$ colors. 
For the other two networks,
\emph{CollegeMsg} and \emph{soc-sign-bitcoinotc} networks, 
CTPLS found $38$ and $40$ colors out of $50$, respectively.
% other vertices with
% the same color. 
For networks with $50$ colors, CTPLS 
found the same number of colors in both
YES-OP and NO-OP networks. 

The experiments on real-world datasets confirm that CTPLS is able to produce suboptimal results even for networks with $50$ colors.
For the networks with $30$ colors, CTPLS found solutions with at least  $63\%$ colors compared to the optimum
(\emph{soc-sign-bitcoinalpha})
and in one case an optimal solution. For the networks with larger number of colors ($50$ colors) CTPLS found solutions with at least  $72\%$ and at most $98\%$
colors compared to the optimum.
Except for (\emph{soc-sign-bitcoinalpha}),
the quality of solution
returned by CTPLS starts slowly to degrade
going from $30$ colors to  $50$ colors.
However, this deterioration is less pronounced than in synthetic datasets.
% with $50$ colors.

As for the running time, CTPLS is able to find a solution of \ColorPath{}
in reasonable time, even for a set of $50$ colors
(notice that this value is larger than what has been considered in \cite{DBLP:journals/bigdata/ThejaswiGL20}). 
The running time varies considerably depending on the size of the temporal network and, in particular, on the length 
of the time domain.
CTPLS indeed has highest running time
on \emph{CollegeMsg} and 
\emph{email-Eu-core-temporal} whose time domain
consists respectively of $207880$
and $58911$ timestamps.
On the other hand, 
CTPLS requires at most
$0.51$ minutes 
%and $0.22$
%minutes 
on \emph{soc-sign-bitcoinalpha} (NP-OP, $50$ colors), which has the smallest time domain (1647 timestamps).

% has highest running time, with the largest  timestamp) and highest running time, then  with second highest time domain () and second highest running time. \emph{soc-sign-bitcoinotc} has less time domain (35445 timestamp) respect to the \emph{email-Eu-core-temporal} and \emph{CollegeMsg} and it is faster than these two networks, while it is slower than 

 \begin{table}[htb]
 \centering % centering table
 \caption {Performance of CTPLS on real datasets. The value of the time (in minutes) and the value of return solution (path) for the \ColorPath{} problem is reported for two different color set (30 and 50).}

 \begin{tabular}{|l||c|c||c|c||} % creating  columns
 \hline 
 \multicolumn{1}{|c|}{}&\multicolumn{2}{|c|}{color 30}&\multicolumn{2}{|c|}{color 50} \\
\hline
NO-OP&path& time&path& time\\ 
 \hline
CollegeMsg&27&144.71&38&27.91 \\ 
email-Eu-core-temporal&30&52.60&49&129.05 \\
soc-sign-bitcoinalpha&20&0.34&36&0.51 \\
soc-sign-bitcoinotc&25&10.98&40&10.35\\
 \hline

\end{tabular}

 \begin{tabular}{|l||c|c||c|c||} % creating  columns
 \hline 
 \multicolumn{1}{|c|}{}&\multicolumn{2}{|c|}{color 30}&\multicolumn{2}{|c|}{color 50} \\
\hline
YES-OP&path& time&path& time\\ 
 \hline
CollegeMsg&27&149.68&38&29.09 \\ 
email-Eu-core-temporal&-&-&49&148.33 \\ 
soc-sign-bitcoinalpha&19&0.16&36&0.22 \\
soc-sign-bitcoinotc&27&6.34&40&9.82\\
 \hline

\end{tabular}

 \label{table:real}
 \end{table}

\section{Conclusion}
In this paper, we have introduced a problem called \ColorPath{} for finding a colorful temporal path of maximum length in a vertex-colored temporal graph.
% The temporal graph is vertex-colored
% and the \ColorPath{} looks for a path whose vertices have distinct colors and include 
% the maximum number of color. 
We have studied the approximation
complexity of the problem and
we have provided an inapproximability lower bound. 
Then we have presented a heuristic (CTPLS)
based on a greedy preliminary step and local search. We have provided an experimental evaluation, both on synthetic and real-world graphs. 
The experimental results on synthetic datasets
have shown that CTPLS returns near optimal solutions for a set of 10 colors, while the performance degrades when the number of colors increases. 
%On the synthetic data the heuristic is always fast.
On the real-world datasets, the algorithm in many cases is able to find suboptimal results in reasonable time, even for networks with 50 colors, despite the fact
that \ColorPath{} is hard to approximate.

Future works include the application of CTPLS to larger temporal networks.
It would also be interesting to consider
whether it is possible to apply 
the algebraic approach proposed in \cite{DBLP:journals/bigdata/ThejaswiGL20} to the \ColorPath{} problem and compare its performance
with CTPLS.  

%In this paper, we propose a new  heuristic (CTPLS) for \ColorPath{} problem, a problem recently introduced for finding maximum colorful paths in temporal graphs. The temporal graph is vertex-colored and the CTPLS asks for a path whose vertices have distinct colors and include the maximum number of color. 
%In particular, we presented the approximation complexity of the problem and we provided an inapproximability lower bound. 
%Then we presented efficient CTPLS based on local search, and an experimental evaluation of our heuristic, both on synthetic and real-world graphs. The experimental results show that CTPLS performed well on both synthetic and real-world networks and is able to find suboptimal results in reasonable time even for networks with 50 colors despite the problems is NP-hard.

\bibliographystyle{splncs03}
\bibliography{Biblio.bib}
%%%% App

\newpage

\section*{Appendix}

\subsection*{Proof of Lemma \ref{lem:IndColoring1}}

\setcounter{lemma}{0}

\begin{lemma}
\label{appendix:lem:IndColoring1}
Let $G_I=(V_I, E_I)$ be an instance of \IS{} and let $G=(V, \TDom, E, c)$ be the corresponding instance of \ColorPath{}. Then:
\begin{enumerate}

\item Each temporal path $p(V_i)$, with $1 \leq i \leq n+1$, is colorful

\item The vertices in temporal paths $p(V_i)$, $p(V_j)$, with 
$1 \leq i < j \leq n$ and $\{ v_i,v_j\} \notin E_I$, 
have different colors.
\end{enumerate}
\end{lemma}
\begin{proof}
1. The property follows from the fact that
each vertex of $V_i$, $1 \leq i \leq n+1$, is associated with a distinct color. 

2. By definition of coloring $c$, since $\{ v_i,v_j\} \notin E_I$, it follows that 
$c(v_{i,j}) = a_i^j$ and $c(v_{j,i}) = a_j^i$
and, by construction, $a_i^j \neq a_j^i$.
Since by construction all the other vertices of $p(V_i)$ and $p(V_j)$
have different colors, it follows that the lemma holds.
\qed
\end{proof}

\subsection*{Proof of Lemma \ref{lem:IndColoring3}}

\setcounter{lemma}{2}

\begin{lemma}
\label{appendix:lem:IndColoring3}
Let $G_I=(V_I, E_I)$ be an instance of \IS{} and let $G=(V, \TDom, E, c)$ be the corresponding
instance of \ColorPath{}. Given 
a solution of \ColorPath{} of length
$(q+1)(n+1)$, we can construct in polynomial time
an independent set
of $G_I$ of size at least $q$.
\end{lemma}
\begin{proof}
Given a colorful temporal path $p$ in $G$ of
length $(q+1) (n+1)$, 
consider the temporal paths $p(V_i)$, with $1 \leq i \leq n+1$, in $p$. We claim that the last
vertex of $p$ is $v_{n+1,n}$.

If the last vertex of $p$ is $v_{n+1,z}$,
for some $z$ with $1 \leq z < n$, we can compute a path $p'$ such that $|p'| > |p|$ by adding the temporal path that starts from vertex $v_{n+1,z}$ and ends in $v_{n+1,n}$. By construction, $p'$ is a temporal path, as this modification does not violate the time constraint. Furthermore, $p'$ is colorful, since $p$ is colorful and each vertex $v_{n+1,j}$, $1 \leq j \leq n$ is assigned a color distinct from the other vertices of $G$. Thus $p'$ is colorful.

If the last vertex of $p$ is $v_{b,n}$,  similarly as the previous case, we can compute a temporal colorful path $p'$, with $|p'|>|p|$, by adding the path $p(V_{n+1})$ to $p$.

If the last vertex of $p$ is $v_{b,z}$, for some $b$ with $1 \leq b \leq n$ and some $z$ with $1 \leq z < n$, we can compute a path $p'$ with $|p'| > |p|$, as follows: (1) we remove from $p$ the temporal path connecting vertices $v_{b,j}$, with $1 \leq j \leq z$, and (2) we add the colorful path $p(V_{n+1})$. Notice that $|p'| \geq |p|$, since $p(V_{n+1})$ contains $n+1$ vertices. Notice that it is always possible to add path $p(V_{n+1})$, since there is a temporal edge $\{v_{a,n}, v_{n+1,0}, t\}$, where the last temporal edge in path $p(V_a)$ is $\{v_{a,n-1}, v_{a,n}, t-1\}$, hence the modification does not violate the time constraint. Since each vertex $v_{n+1,j}$, $1 \leq j \leq n$, has color distinct from the other vertices of $G$, it follows that $p'$ is colorful.

We claim now that $p'$ contains at least $q$ temporal paths  $p(V_i)$, $1 \leq i \leq n$. Assume that this is not the case, it follows that by construction the temporal colorful path $p'$ contains less than $q$ paths $p(V_i)$, with $1 \leq i \leq n$ and path $p(V_{n+1})$. Since by construction $|p(V_j)| = n+1$, with $1 \leq j \leq n+1$, it follows that $|p'| < (q +1)(n+1)$.

Now, consider two paths $p(V_i)$, $p(V_j)$ in $p'$, with $1 \leq i < j \leq n$. Since $p'$ is colorful, it follows that the vertices of $p(V_i)$ and $p(V_j)$  are associated with different colors Then $\{ v_i, v_j\} \notin E_I$, otherwise the two vertices $v_{i,j}$ and $v_{j,i}$ in $p(V_i)$, $p(V_j)$, respectively, are both assigned  the same color $c_{i,j}$. It follows that we can define an independent set $I$ of $G_I$ as follows:
\[
I = \{ v_i: p(V_i) \text{ is a path of } p'\} .
\]
Notice that since $|p'| > |p| \geq (q+1)(n+1)$, it follows that $I$ is an independent set of size at least $q$, thus concluding the proof.
\end{proof}

\subsection*{Proof of Theorem \ref{teo:inapprox}}

\setcounter{theorem}{0}

\begin{theorem}
\label{appendix:teo:inapprox}
\ColorPath{} is not approximable within factor $O(|V|^{1/2- \varepsilon})$ unless P = NP.
\end{theorem}
\begin{proof}
We show now that the reduction we have described is indeed an approximation preserving reduction. Denote the value of an optimal solution of \ColorPath{} (\IS{}, respectively) by $OPT(CPTG)$ ($OPT(IS)$, respectively); denote the value of an approximate solution of \ColorPath{} (\IS{}, respectively) by $APX(CPTG)$ ($APX(IS)$, respectively). Next, consider the approximation factor of \ColorPath{}, that is
\[
\frac{OPT(CPTG)}{APX(CPTG)}.
\]
By Lemma \ref{lem:IndColoring2}, it follows that $OPT(CPTG) \geq (n+1)\cdot (OPT(IS) + 1)$. Thus
\[
\frac{OPT(CPTG)}{APX(CPTG)}
\geq \frac{(n+1) \cdot (OPT(IS)+1)}{APX(CPTG)}.
\]

By Lemma \ref{lem:IndColoring3}, given an approximated solution of $\ColorPath{}$ of size $(n+1)\cdot (q+1)$, we can compute in polynomial time a solution of \IS{} of size at least $q$. It follows that $APX(CPTG) \leq (n+1)\cdot (APX(IS) + 1)$. Thus
\[
\frac{OPT(CPTG)}{APX(CPTG)}
\geq \frac{(n+1) \cdot (OPT(IS)+1)}{APX(CPTG)}
\geq \frac{(n+1) \cdot (OPT(IS)+1)}{(n+1) \cdot( APX(IS)+1)}.
\]

Since we can assume that $APX(IS) \geq 1$ it follows that
\[
\frac{OPT(CPTG)}{APX(CPTG)}
\geq \frac{OPT(IS)+1}{APX(IS)+1}
\geq \frac{OPT(IS)}{2 APX(IS)}.
\]

Since $\IS$ is not approximable within factor $O(n^{1-\varepsilon})$, for any $\varepsilon > 0$ unless P = NP \cite{DBLP:conf/stoc/Zuckerman06}, it follows that 
\[
\frac{OPT(CPTG)}{APX(CPTG)}
\geq \frac{OPT(IS)}{2APX(IS)} = O(n^{1 -\varepsilon}).
\]
By construction, $|V| = (n+1)^2$, hence we have that

\[
\frac{OPT(CPTG)}{APX(CPTG)}
\geq  
O(n^{1 -\varepsilon}) =
O(|V|^{\frac{1}{2} -\varepsilon}-1) =
O(|V|^{\frac{1}{2} -\varepsilon}),
\]
thus concluding the proof.
\end{proof}

\end{document}